\documentclass[%
 reprint,
 superscriptaddress,
 dvipdfmx,
 amsmath,amssymb,
 aps,
pra,
]{revtex4-1}

\usepackage{graphicx}
\usepackage{dcolumn}
\usepackage{bm}
\usepackage{booktabs}
\usepackage{color}

\begin{document}


\title{Neutron quadrupole transition strength in $^{10}$C
deduced from the $^{10}$C$(\alpha,\alpha')$ measurement with the MAIKo active target}

\author{T.~Furuno}
\affiliation{Research Center for Nuclear Physics (RCNP), Osaka University, 
  Ibaraki, Osaka 567-0047, Japan}

\author{T.~Kawabata}
\affiliation{Department of Physics, Osaka University, 
  Toyonaka, Osaka 560-0043, Japan}

\author{S.~Adachi}
\affiliation{Department of Physics, Kyushu University, 
  Fukuoka, Fukuoka 819-0395, Japan}

\author{Y.~Ayyad}
\affiliation{Facility for Rare Isotope Beams (FRIB), 
  Michigan State University, East Lansing, Michigan 48824, USA}

\author{Y.~Kanada-En'yo}
\affiliation{Department of Physics, Kyoto University, 
  Sakyo, Kyoto 606-8502, Japan}

\author{Y.~Fujikawa}
\affiliation{Department of Physics, Kyoto University, 
  Sakyo, Kyoto 606-8502, Japan}

\author{K.~Inaba}
\affiliation{Department of Physics, Kyoto University, 
  Sakyo, Kyoto 606-8502, Japan}

\author{M.~Murata}
\altaffiliation[Present Address: ]{Research Center for Nuclear Physics (RCNP), 
  Osaka University, Ibaraki, Osaka 567-0047, Japan}
\affiliation{Department of Physics, Kyoto University, 
  Sakyo, Kyoto 606-8502, Japan}

\author{H.~J.~Ong}
\affiliation{Research Center for Nuclear Physics (RCNP), Osaka University, 
  Ibaraki, Osaka 567-0047, Japan}

\author{M.~Sferrazza}
\affiliation{D\'epartment de Physique, Universit\'e Libre de Bruxelles, 
  Bruxelles 1050, Belgium}

\author{Y.~Takahashi}
\affiliation{Department of Physics, Kyoto University, 
  Sakyo, Kyoto 606-8502, Japan}

\author{T.~Takeda}
\affiliation{Department of Physics, Kyoto University, 
  Sakyo, Kyoto 606-8502, Japan}

\author{I.~Tanihata}
\affiliation{Research Center for Nuclear Physics (RCNP), Osaka University, 
  Ibaraki, Osaka 567-0047, Japan}
\affiliation{International Research Center for Nuclei and Particles in the Cosmos, 
  and School of Physics and Nuclear Energy Engineering, Beihang University, 
  Beijing 100191, China}

\author{D.~T.~Tran}
\affiliation{Research Center for Nuclear Physics (RCNP), Osaka University, 
  Ibaraki, Osaka 567-0047, Japan}
\affiliation{Institute of Physics, Vietnam Academy of Science and Technology, 
  BaDinh, Hanoi 100000, Vietnam}

\author{M.~Tsumura}
\affiliation{Department of Physics, Kyoto University, 
  Sakyo, Kyoto 606-8502, Japan}

\date{\today}

\begin{abstract}
  Elastic and inelastic alpha scatterings on $^{10}$C were measured 
  using
  a 68-MeV/u radioactive $^{10}$C beam incident on the recently developed
  MAIKo active target system.
  The phenomenological effective $\alpha$-$N$ interaction and 
  the point-nucleon density distribution in the ground state were 
  determined from the elastic scattering data.
  The cross sections of the inelastic alpha scattering were 
  calculated using this interaction and density distribution
  and were compared with the experiment to determine the neutron 
  quadrupole transition matrix element $M_{n}$ between the ground state and the 
  $2_{1}^{+}$ state at $E_{x} = 3.35$ MeV in $^{10}$C.
  The deduced neutron transition matrix element is
  $M_{n} = 6.9\, \pm0.7\, \mathrm{(fit)}\, \pm1.2\, \mathrm{(sys)}$ fm$^{2}$.
  The ratio of the neutron transition strength to proton transition strength 
  was determined as
  $M_{n}/M_{p} = 1.05\, \pm0.11\, \mathrm{(fit)}\, \pm0.17\, \mathrm{(sys)}$,
  which indicates that the quadrupole transition between the ground state and
  the $2_{1}^{+}$ state in $^{10}$C is less neutron dominant
  compared to that in $^{16}$C.
\end{abstract}

\maketitle

\section{\label{sec-intro}Introduction}


Quadrupole transitions between nuclear states provide 
valuable insight into nuclear structure.
Their strengths are key benchmarks in testing theoretical models.
In particular, the quadrupole transition strengths from the ground ($0_{1}^{+}$) state
to the $2_{1}^{+}$ state in even-even nuclei reflect nuclear shell structures
\cite{PhysRevLett.42.425,bernstein_plb,bernstein_com,PhysRevC.46.1811,PhysRevC.46.164}.
The quadrupole transitions are described as rearrangements of particle-hole
configurations in the valence shells under the framework of the nuclear shell model.
When the neutron (proton) shell is closed, 
the transition of the neutron (proton) is suppressed remarkably 
because the intra-shell excitation in the closed shell is forbidden. 
As a consequence, the ratio of the neutron and proton transition strengths
deviates from unity.


The neutron (proton) transition matrix element from the $0_{1}^{+}$
state to the $2_{1}^{+}$ state is defined as follows:
\begin{equation}
  M_{n(p)}=\langle 2_{1}^{+} || \sum_{n(p)} r^{2}Y_{2}|| 0_{1}^{+} \rangle.
  \label{matrix-element}
\end{equation}
Here, $Y_{2}$ represent the spherical functions for $L=2$.
Considering the proton as a point particle,
$M_{p}$ can be related to
the reduced electric quadrupole transition rate 
$B(E2;0_{1}^{+} \to 2_{1}^{+})$ by the following relation:
\begin{equation}
  B(E2;0_{1}^{+}\to2_{1}^{+})=e^2|M_{p}|^2.
  \label{eq_be2}
\end{equation}
The relationship between excitation ($0_{1}^{+} \to 2_{1}^{+}$) and 
deexcitation ($2_{1}^{+} \to 0_{1}^{+}$) 
reduced transition rates can be given by a simple equation
\begin{equation}
\begin{split}
  B(E2;0_{1}^{+} \to 2_{1}^{+}) =& \frac{2J'+1}{2J+1} B(E2;2_{1}^{+} \to 0_{1}^{+})\\
  =& 5 B(E2;2_{1}^{+} \to 0_{1}^{+}),
\end{split}
\end{equation}
where $J$ and $J'$ are the spins of the ground and excited states.

There is no direct way to determine $M_{n}$ since there exists no probe 
that is sensitive only to neutrons. 
To determine $M_{n}$ of a nucleus, one can either adopt $M_{p}$ of 
its mirror nucleus assuming charge symmetry,
or disentangle $M_{n}$ from inelastic scattering cross sections using a 
hadronic probe such as a proton or alpha particle, 
incorporating $M_{p}$ obtained from the $B(E2;0_{1}^{+} \to 2_{1}^{+})$ value.


With the progress on techniques for providing radioactive isotope (RI) beams
over the past few decades,
numerous efforts have been made to deduce transition matrix elements in
unstable nuclei
\cite{riley_prl_1999,Jewell1999,Iwasaki2000,Khan2001,cottle_prl_2002,Elekes2004,imai_16c,10cp,riley_prc_2005,yamada_2005,ong_16c,PhysRevLett.96.012501,Campbell2007,PhysRevLett.100.152501,ong_18c,PhysRevC.79.011302,PhysRevLett.107.102501,PhysRevC.86.044329,PhysRevC.86.011303,togano_prl_2012,michimasa_prc_2014,PhysRevC.90.011305,Corsi2015,PhysRevC.97.044315,PhysRevC.100.044312}.
The highlights are the discovery of the
enhanced $M_{n}/M_{p}$ ratio and the
suppression of $M_{p}$ in neutron-rich carbon isotopes.
%
The $M_{n}/M_{p}$ ratio in $^{16}$C is as large as 3 
\cite{Elekes2004,ong_16c,ong_18c},
which is much larger than $N/Z=1.7$.
The reduced electric quadrupole transition rates 
$B(E2;2_{1}^{+} \to 0_{1}^{+})$ in 
$^{16}$C \cite{imai_16c,PhysRevLett.100.152501,PhysRevC.86.044329,ong_18c},
$^{18}$C \cite{ong_18c,PhysRevC.86.011303},
and $^{20}$C \cite{PhysRevLett.107.102501}
are small, i.e., approximately
1.1--2.3 Weisskopf units (W.u.).

In Ref. \cite{z6magic}, the large $M_{n}/M_{p}$ ratio or quenching of $M_{p}$
in neutron-rich carbon isotopes was attributed to the subshell closure at $Z=6$.
It is of interest to see whether the large $M_{n}/M_{p}$ ratio is observed
in a proton-rich carbon isotope as well, for example, $^{10}$C.


The $B(E2;2_{1}^{+} \to 0_{1}^{+})$ value in $^{10}$C had been known to be
$9.6\pm1.6$ W.u. \cite{be2_10c_old}.
This value is not small as compared to those in the neutron-rich carbon
isotopes.
From an inelastic proton scattering experiment \cite{10cp},
the $M_{n}/M_{p}$ ratio in $^{10}$C was measured to be $0.70\pm0.08$,
which is not so large compared to that in neutron-rich $^{16}$C.
However, a more recent lifetime measurement on the $2_{1}^{+}$ state in $^{10}$C
reported a smaller value of 
$B(E2;2_{1}^{+} \to 0_{1}^{+})=6.9\pm0.2$ W.u. \cite{10c_be2}
with uncertainty much smaller in comparison with
the previous measurement.
Therefore, it is important to revisit the $M_{n}/M_{p}$ ratio 
in $^{10}$C.
To determine $M_{n}$ in $^{10}$C, instead of the inelastic proton scattering,
we utilized the inelastic alpha scattering, which is an isoscalar probe.

In order to determine the neutron matrix element $M_{n}$ from the
cross section of the inelastic hadron scattering,
the Bernstein prescription:
$d\sigma / d\Omega \propto |b_{n}^{F}M_{n} + b_{p}^{F}M_{p}|^{2}$
is often used \cite{bernstein_plb}.
Here, $b_{n}^{F}$ and $b_{p}^{F}$ are external-field interaction strengths
that reflect the effective interaction between an incident particle
and a proton or neutron in nuclei.
In the case of $(p,p')$ scattering, 
the ratio $b_{n}/b_{p}$ is phenomenologically determined.
However, this ratio has a strong energy dependence;
it varies from 3 to 0.83 
in the incident-energy range of 10--1000 MeV \cite{bernstein_com},
and it also depends on the nucleus.
This dependence causes a significant model ambiguity in the determination of 
the $M_{n}/M_{p}$ ratio from $(p,p')$ scattering.
On the other hand, in $(\alpha, \alpha ')$ scattering,
the $b_{n}^{F}/b_{p}^{F}$ ratio is always unity because 
of the zero isospin of an alpha particle.
Therefore, $(\alpha, \alpha ')$ scattering is more suited to deduce
$M_{n}$ than $(p,p')$ scattering.

To deduce $M_{n}$ from the measured cross section,
distorted-wave Born-approximation (DWBA) calculations should be performed.
Recently, the alpha inelastic scattering off the self-conjugate even-even nuclei
from $^{12}$C to $^{40}$Ca was systematically measured \cite{adachi}.
The DWBA calculations using single-folding model potentials reasonably
reproduce the measured cross sections,
once the effective $\alpha$-$N$ interaction
is determined to reproduce the elastic alpha scattering.
Therefore, the cross sections of the elastic as well as 
inelastic alpha scattering should be measured.

Since $^{10}$C is unstable, the measurement must be done in the inverse
kinematic condition.
One of the best methods to measure the cross sections of the
elastic and inelastic scatterings
in a single experiment is missing mass spectroscopy,
in which the excitation energies of incoming nuclei are determined 
from energies and angles of recoil particles.
In fact, the previous $(p,p)$ and $(p,p')$ 
measurements were
performed using missing mass spectroscopy \cite{10cp}.
However, it is not easy to apply missing mass spectroscopy to
alpha inelastic scattering,
especially at low momentum transfer of $q\sim0.5$ fm$^{-1}$
where the differential cross section for the $2_{1}^{+}$
state becomes maximum.
Because the energies of the recoil alpha particles are only 
$E_{\alpha}\sim1$ MeV at $q\sim0.5$ fm$^{-1}$, 
it is almost impossible to detect low-energy particles
by a conventional experimental setup.

To overcome the challenge of detecting low-energy recoil alpha particles,
we recently developed the MAIKo active target \cite{maiko}.
This system consists of a time projection chamber (TPC)
which is a gaseous detector with three-dimensional reconstruction capability
for charged particle trajectories.
In the active target mode, the detection medium gas of the TPC
is also used as the target gas.
The detection of low-energy recoil particles becomes possible
because the reaction occurs inside the sensitive volume of the TPC

In this paper, we report the measurements of the cross sections for the
$\alpha+^{10}$C elastic and inelastic scatterings at 
an incident energy of 68 MeV/u.
The present experiment is the first attempt to deduce the $M_{n}$ value
in unstable nuclei from $(\alpha, \alpha ')$ scattering
and the first experiment using the MAIKo active target.
The cross sections were measured
at $\theta_{\mathrm{c.m.}}=4$--15$^{\circ}$, which corresponds to
the momentum transfer of $q=0.4$--$1.4$ fm$^{-1}$.
The neutron transition matrix element $M_{n}$ in $^{10}$C
was determined and the $M_{n}/M_{p}$ ratio is discussed.

\section{\label{sec-exp}Experiment}
\begin{figure*}[tb]
\includegraphics[width=160mm]{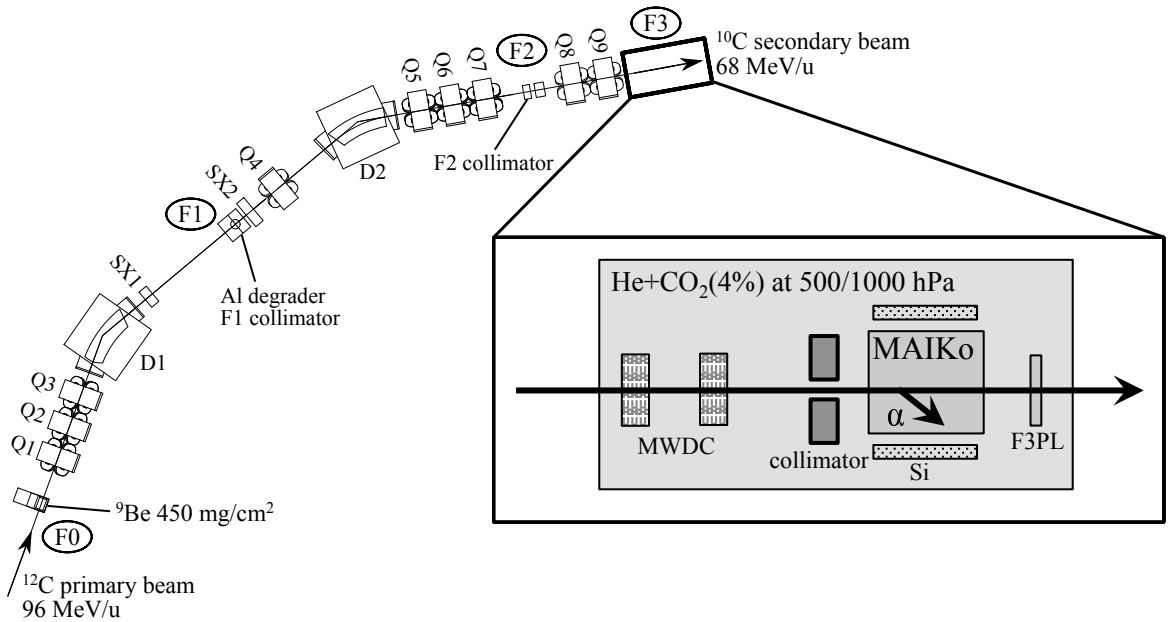}
\caption{Schematic view of the EN course and the beamline detectors.
  The inset shows the details of the detector setup at F3.}
\label{beamline}
\end{figure*}

The measurement was carried out at the cyclotron facility of
Research Center for Nuclear Physics (RCNP), Osaka University.
A $^{10}$C secondary beam was produced via projectile fragmentation using
a $^{12}$C$^{6+}$ primary beam 
at 96 MeV/u accelerated by the azimuthally varying field cyclotron 
and the ring cyclotron.
The primary beam with intensities ranging from 50 to 100 pnA 
was transported 
to the exotic nuclei (EN) beamline shown in Fig. \ref{beamline}, 
and incident on a 
450-mg/cm$^{2}$-thick $^{9}$Be production target at the F0 focal plane.
$^{10}$C particles were separated from other fragments 
using the fragment separator at the EN beamline \cite{en1,en2,en3}
by setting the appropriate magnetic rigidities in two dipole magnets (D1 and D2).
To improve isotope separation, a 2-mm-thick aluminum degrader was placed at the
first focal plane (F1).
The $^{10}$C beam was angular focused at the second focal plane (F2),
which is a charge-mass dispersive focal plane,
and further selected using a pair of collimators at F2.
The selected $^{10}$C beam was later transported to the third focal plane (F3),
and injected into the MAIKo active target.
During the tuning of the secondary beam, the purity of $^{10}$C was measured
from the correlation between the energy loss of the beam particles in a Si
detector before MAIKo and the time of flight.
The time of flight was measured between the F0 target and the Si detector
using the radio-frequency signal from the accelerator.
Because the purity of $^{10}$C was as high as 96$\%$ before MAIKo, 
event-by-event beam particle identification (PID)
was not necessary for the present work.

The details of the detector setup at F3 are shown in the inset of Fig. \ref{beamline}.
The intensity of the beam was measured with a 
1-mm-thick plastic scintillator (F3PL) placed downstream of MAIKo at F3.
The typical beam intensity was 70,000 counts per second (cps).
We placed two multi-wire drift chambers (MWDCs) before MAIKo
for monitoring the profile of the incident beam.
The distance between the two MWDCs was 600 mm,
and the distance from the downstream MWDC to the entrance of MAIKo was 733 mm.
The $^{10}$C beam was horizontally collimated to $\pm10$ mm
by 10-mm-thick tungsten collimators before MAIKo.
The angular spread of the $^{10}$C beam was 15 mrad in the
horizontal direction and 6 mrad in the vertical direction.
The spot size of the $^{10}$C beam at the entrance of MAIKo was
5 mm both in horizontal and vertical directions.
The average energy of the $^{10}$C beam was 68 MeV/u before the F3PL
with an energy spread of 1$\%$.

\begin{figure}[tb]
\includegraphics[width=85mm]{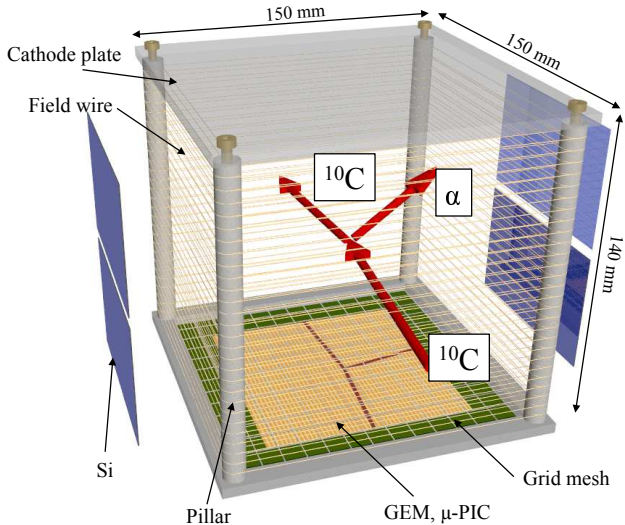}
\caption{Schematic view of the MAIKo active target.}
\label{fig_maiko}
\end{figure}

Figure \ref{fig_maiko} shows the schematic view of the MAIKo active target.
The TPC field cage has dimensions of 150 $\times$ 150 $\times$ 140 mm$^{3}$.
The angle from the beam axis and the kinetic energy of the
recoil alpha particle are measured to determine the excitation energy
of $^{10}$C and the scattering angle in the center-of-mass (c.m.) frame.
The recoil angle is determined from the reconstructed trajectory of the
recoil alpha particle.
The kinetic energy is determined from the length of the trajectory
when the recoil alpha particles stop in the sensitive volume of the TPC.
High-energy recoil alpha particles that escape from the TPC sensitive
volume are detected by four silicon detectors placed outside
the TPC.
Each silicon detector has a sensitive area of 90 $\times$ 60 mm$^{2}$,
and a thickness of 500 $\mu$m.
When a recoil alpha particle stops between the silicon detectors
and the TPC, the recoil energy cannot be determined.
Because the insensitive energy region depends on the 
gas pressure, the TPC was operated at two different gas pressures.

The TPC was operated with a He $+$ CO$_{2}(4\%)$ mixture gas
at 500 and 1000 hPa.
When the gas pressure is 500 hPa, alpha particles with kinetic energies
$E_{\alpha}<1$ MeV stop in the sensitive volume of the TPC, while
alpha particles with $E_{\alpha}>4$ MeV reach the silicon detectors.
When the gas pressure is at 1000 hPa, alpha particles with
$E_{\alpha}<3$ MeV stop inside the TPC.

The TPC shared the same gas chamber with the MWDCs, 
and thus, the TPC and MWDCs were operated using the same detection gas.
The He and CO$_{2}$ gasses were individually supplied using two mass flow
controllers to keep the mixing ratio and the flow rate constant.
The total flow rate was set at 100 cm$^{3}$/min.
The pressure and the temperature inside the chamber were monitored using
a diaphragm gauge and a Pt-100 thermometer, respectively.
The gas density was determined from the pressure and temperature.
The mixture gas was exhausted from the chamber using a scroll pump.
A piezo valve was installed between the chamber and the pump.
The aperture of the valve was automatically controlled to keep the gas density
constant according to the measured pressure and temperature.
The density fluctuation was within $\pm$0.2$\%$ throughout the measurement.

The red arrows in Fig. \ref{fig_maiko} present trajectories of
an incident $^{10}$C, a scattered $^{10}$C, and a recoil alpha particle.
The $^{10}$C nuclei and recoil alpha particles ionize the gas molecules along their
trajectories.
The ionized electrons
drift vertically along the electric field formed by the TPC field cage.
The electric field was formed by applying negative high voltages on
the stainless-steel cathode plate and the nickel grid mesh.
The cathode plate and the grid mesh were kept at a distance of 140 mm by 
four pillars made of Macor ceramic.
Field wires made of beryllium copper were doubly wound around the pillars
with 5-mm intervals to make the electric field uniform.
The strength of the electric field was chosen such that the
electron-drift velocity was 1.7 cm/$\mu$s for both 500 and 1000 hPa pressures. 
The electron-drift velocity was measured using collimated alpha particles
from a $^{241}$Am source.
The above drift velocity is high enough to collect the electrons from the
full active volume, within the time window of 10.24 $\mu$s.

After the drifted electrons pass through the grid mesh,
they are multiplied first through a gas electron
multiplier (GEM) and then by a micro-pixel chamber ($\mu$-PIC) \cite{ochi-mupic}.
The total gas gain of the GEM and the $\mu$-PIC was 
measured to be about 870 in operation at 500 hPa.
The $\mu$-PIC was also used to measure the position of the drifted electrons.
The $\mu$-PIC has a sensitive area of 102.4 $\times$ 102.4 mm$^{2}$.
It consists of 256 anode strips and 256 cathode strips which are arranged
orthogonally.
These strips are fabricated at 400-$\mu$m intervals.
The signals induced by the electron avalanche are read out through the
anode and cathode strips which provide the 
two-dimensional information of the particle trajectories.
The vertical position of the trajectories are determined from the electron
drift time multiplied by the drift velocity.

The analog signals from the anode and cathode strips are pre-amplified,
shaped, and discriminated with dedicated readout boards \cite{iwaki}.
The discriminators give output, a high or low-level signal, by comparing
the pulse height of the shaped analog signal with a threshold voltage.
The output of the discriminators is synchronized with a 100-MHz clock.
When a trigger signal is provided to the readout boards, 
the status of the discriminators at every 10 ns 
is recorded as a function of the
clock number for a time window of 10.24 $\mu$s.
The summed pulse shapes of the adjacent 32 strips (12.8 mm width)
are also recorded by 25-MHz flash analog-to-digital converters (FADCs).

\begin{figure}[tb]
\includegraphics[width=85mm]{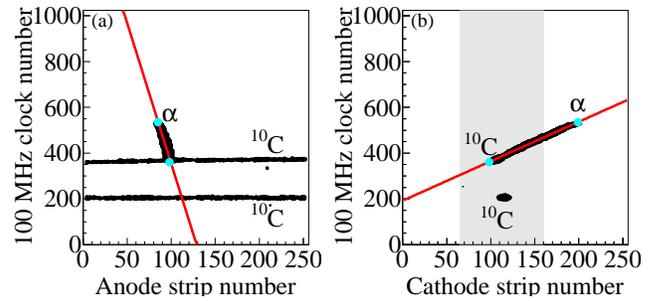}
\caption{
  Example of (a) anode and (b) cathode images acquired in an $\alpha+^{10}$C event.
  The reconstructed trajectories of the recoil particle are drawn with
  the solid red lines.
  The vertex and the track endpoints are shown by the cyan circles.
  The shaded area in the cathode image was excluded from the trigger condition.
}
\label{scatrack}
\end{figure}

\begin{figure}[tb]
\includegraphics[width=85mm]{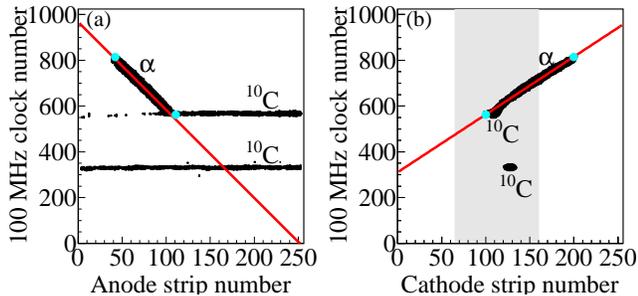}
\caption{
  Same as Fig. \ref{scatrack}, 
  but track images in an inelastic scattering event 
  exciting a highly excited state
  above the proton decay threshold at $E_{x}=3.82$ MeV.
  The trajectories of two protons and two alpha particles from
  the breakup of $^{10}$C are thin because of the small energy loss.
}
\label{scatrack2}
\end{figure}

The data recorded by the discriminators are equivalent to two
black-and-white images with $256\times1024$ pixels.
Each image presents particle trajectories projected onto the plane
perpendicular to the anode or cathode strips.
Examples of the track images of $\alpha+^{10}$C events are
shown in Figs. \ref{scatrack} and \ref{scatrack2}.
In Fig. \ref{scatrack}, another $^{10}$C entering the sensitive volume of MAIKo
about 150 clocks ($=1.5$ $\mu$s) before the scattering event was accidentally
recorded.
The $^{10}$C beam trajectories appear as horizontal loci in the anode images
because the anode strips are perpendicular to the beam axis.
On the other hand, they appear as elliptical shapes in the cathode image
because the cathode strips are parallel to the beam axis.

The trigger signal for the data acquisition was generated from the 
silicon detectors or cathode strips.
To suppress the triggers due to the beam particles,
the 65th--160th cathode strips were excluded from the trigger.
This trick inhibits the shaded area in Figs. \ref{scatrack}(b) 
and \ref{scatrack2}(b) from triggering data acquisition.
When the beam intensity was 77 kHz and the gas pressure was at 500 hPa, 
the trigger rate was 270 Hz,
and the live time ratio of the data acquisition was 88$\%$.

In addition to the measurement using the $^{10}$C beam,
we also performed a similar measurement using the $^{12}$C primary beam 
to compare the cross section measured by MAIKo
with the previous result measured under the normal kinematic
condition \cite{adachi}.

\section{\label{sec-data}Data reduction}
MAIKo acquired
not only the $\alpha+^{10}$C scattering events but also background events.
The background events were mainly caused by 
$^{10}$C beam particles which invade the cathode trigger region
(beam events).
Scattering from the quenching CO$_2$ gas also caused the background events.
The fraction of the $\alpha+^{10}$C events in the acquired events 
was only of the order of 1$\%$.
Therefore, the $\alpha+^{10}$C events must be correctly distinguished from
the background events.

The $\alpha+^{10}$C events exhibit the following two features.
First, energy losses per unit length of recoil alpha particles
are about 7 times larger than those of $^{10}$C beam particles.
Second, because $^{10}$C is much heavier than an alpha particle,
the scattering hardly deflects $^{10}$C;
however alpha particles recoil at large angles.
Therefore, just one trajectory with a large angle from the horizontal line
(non-horizontal trajectory)
due to a recoil alpha particle should be observed in the anode image,
as seen in Figs. \ref{scatrack}(a) and \ref{scatrack2}(a).
On the other hand, in a beam event, 
the non-horizontal trajectory is not recorded.
In a background event due to the CO$_{2}$ gas,
multiple non-horizontal trajectories are observed.

Considering these features, the analysis of the TPC data was performed with
the following procedures.
\begin{enumerate}
  \renewcommand{\labelenumi}{\roman{enumi})}
   \item 
     Eliminate the hit pixels in the anode and cathode images 
     from the analysis,
     where the corresponding induced charges measured with the FADCs are
     smaller than a certain threshold.
   \item 
     Extract the straight lines in the anode and cathode images 
     using the Hough transformation \cite{hough,hough2}
     as described in Ref. \cite{maiko}.
   \item
     If the number of non-horizontal lines in the anode images is 1,
     this line is regarded as the trajectory of the recoil alpha particles.
     Fit the hit pixels near the trajectory of the recoil particle
     to a straight line for a better track determination.
     The fits are shown by the solid red lines in 
     Figs. \ref{scatrack} (a) and \ref{scatrack2} (a).
   \item 
     Ignore the straight lines in the cathode images shorter than a certain threshold
     to remove the trajectories of unreacted $^{10}$C.
     If the number of the remaining straight lines is 1,
     fit the hit pixels to a straight line as done in anode images.
     The fits are shown by the solid red lines in 
     Figs. \ref{scatrack} (b) and \ref{scatrack2} (b).
   \item 
     Find the vertex and track endpoints in the anode
     and cathode images along the fitted lines.
     These points are indicated with the cyan circles
     in Figs. \ref{scatrack} and \ref{scatrack2}.
   \item 
     From the angles of the straight lines in 
     the anode and cathode images,
     calculate the polar and azimuthal recoil angles,
     assuming the beam axis is parallel to the
     cathode strips.
     The range of the recoil particle is determined 
     from the distance between the vertex and the track endpoints.
\end{enumerate}

We selected the $\alpha+^{10}$C events where the vertex point locates 
between the 33rd--224th strips in the anode image.
Thus, the effective length of MAIKo as a He gas target was 76.8 mm.

\begin{figure}[tb]
\includegraphics[width=85mm]{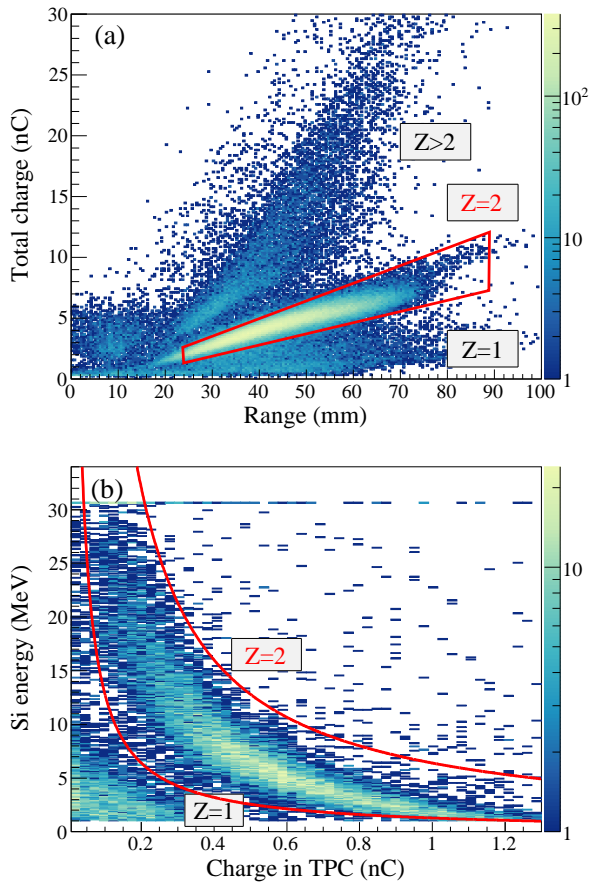}
\caption{
  PID of recoil particles 
  in the measurement at 500 hPa.
  (a) Correlation between the total charge collected by the $\mu$-PIC
  and the range of the recoil particle.
  (b) Correlation between the energy measured with a silicon detector and
  the charge collected with the 32 cathode strips near the silicon detector.
}
\label{pid}
\end{figure}

PID for the recoil particles must be performed.
Ranges of charged particles in the gas are proportional to $E^{2}/(AZ^{2})$,
and total charges collected by the $\mu$-PIC are proportional to kinetic
energies of recoil particles,
if the recoil particles stop in the sensitive volume of the TPC.
Therefore, PID can be performed from the correlation between the total 
charge measured
with the $\mu$-PIC and the range, as shown in Fig. \ref{pid}(a).

If recoil particles escape from the TPC and hit the silicon detectors,
PID can be performed from the correlation between the charge collected by
$\mu$-PIC and the energy measured by the silicon detector, as shown in
Fig. \ref{pid}(b).

In both cases, the loci of the $Z=2$ particles are clearly separated from
the loci of $Z>2$ and $Z=1$ particles.
We selected the $Z=2$ events enclosed by the solid red lines
in Fig. \ref{pid}.
The minimum range for the recoil alpha particle is 25 mm.

If recoil alpha particles stop inside the sensitive volume of the TPC,
recoil energies are calculated from ranges in the gas
using the SRIM code \cite{srim}.
If recoil alpha particles hit the silicon detectors,
recoil energies are calculated using
$E_{\alpha}=E_{\mathrm{Si}}+\Delta E_{\mathrm{gas}}$,
where $E_{\mathrm{Si}}$ is the energy measured with the silicon detector and
$\Delta E_{\mathrm{gas}}$ is the energy loss of the recoil alpha particle 
in the gas.
$\Delta E_{\mathrm{gas}}$ is
calculated by integrating $dE/dx$ along the particle trajectory
between the silicon detector and the vertex position.

\begin{figure}[tb]
\includegraphics[width=78mm]{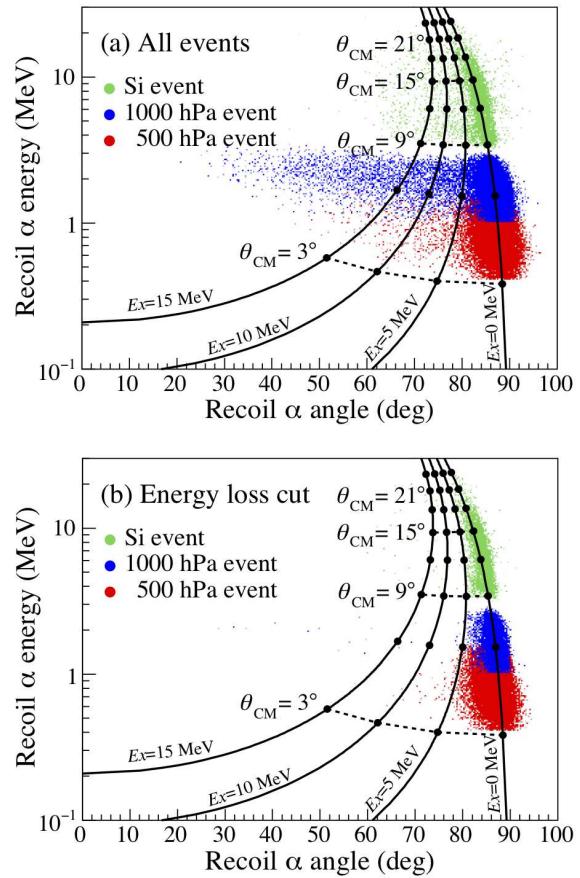}
\vspace{2mm}
\caption{
  Scatter plots of kinematic energies versus angles
  of recoil alpha particles.
  (a): All of the $\alpha+^{10}$C events.
  (b): The elastic and inelastic scattering
  events to low excited states selected by the FADC data.
  The red, blue, and green dots show the ``500 hPa'', ``1000 hPa'', 
  and ``Si'' events, respectively.
  Kinematically calculated energies and angles of recoil alpha particles
  at different excitation energies in $^{10}$C 
  are shown by the solid lines. 
  }
\label{kinema}
\end{figure}

Figure \ref{kinema} shows the reconstructed recoil energy versus the recoil 
angle in the $\alpha+^{10}$C events.
The red and blue dots represent events in which a recoil alpha particle
stopped inside the TPC during the measurements at
500 and 1000 hPa
(denoted as ``500 hPa event'' and ``1000 hPa event''), respectively.
The green dots represent events in which a recoil alpha particle hit
one of the silicon detectors during the measurement at 500 hPa 
(denoted as ``Si event'').
With the present measurement, we successfully lowered the detection 
threshold to 0.5 MeV.
This detection threshold is determined by the minimum range of 25 mm
defined in the PID procedure.
The calculated energies and angles of recoil alpha particles at different 
excitation energies in $^{10}$C are shown with the solid lines.

Excited states in $^{10}$C below the proton emission threshold at
$E_{x}=3.82$ MeV
always decay to the ground state by emitting a $\gamma$-ray.
Because both the incident and scattered particles are $^{10}$C,
the energy loss of $^{10}$C per unit length in the TPC gas after the 
scattering point is almost the same as that before the scattering point.
Therefore,
thicknesses of the observed horizontal trajectories in the anode image look
similar before and after the scattering point, as seen in 
Fig. \ref{scatrack}(a).
On the other hand,
when the excitation energy is above the proton emission threshold,
the excited states immediately decay
into $2p + 2\alpha$ particles.
Since energy loss of the $2p + 2\alpha$ particles is about 1/3 of
the incident $^{10}$C, 
the observed horizontal trajectories after the scattering
point look thinner than before the scattering point 
as seen in Fig. \ref{scatrack2}(a).
Therefore, inelastic scattering to highly excited states at $E_{x}>3.82$ MeV
is easily discriminated from the elastic scattering and inelastic scattering
to low excited states at $E_{x}<3.82$ MeV by using the energy-loss information
obtained from the most downstream channel of the FADC for the
1st--32nd anode strips.
Figure \ref{kinema}(a) includes all of the $\alpha+^{10}$C events,
while Fig. \ref{kinema}(b) includes the elastic and inelastic scattering
events to low excited states selected by the FADC data.
Using energy loss information, only low-lying states below the particle decay 
threshold were successfully selected.
In the present work, we focus on the low-excitation energy events.

\begin{figure}[tb]
\includegraphics[width=85mm]{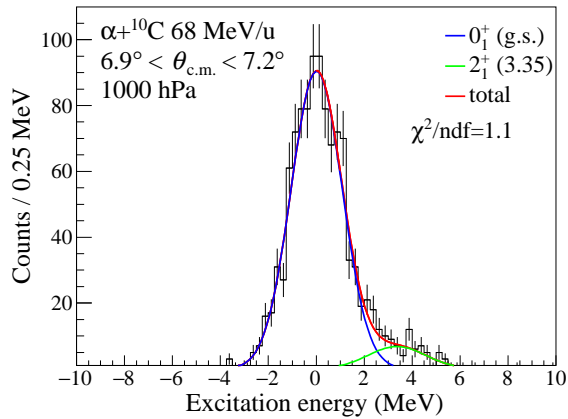}
\caption{
  Excitation energy spectrum in the $\alpha+^{10}$C 
  scattering at $E=68$ MeV/u and
  $6.9^{\circ} < \theta_{\mathrm{c.m.}} < 7.2^{\circ}$.
  The solid lines represent the fit result of
  0$^{+}_{1}$ (blue), 2$^{+}_{1}$ (green), and sum of the two states (red),
  respectively.
}
\label{exfit}
\end{figure}

Figure \ref{exfit} shows the excitation-energy spectrum in the 
$\alpha+^{10}$C scattering at $E=68$ MeV/u and 
$6.9^{\circ} < \theta_{\mathrm{c.m.}} < 7.2^{\circ}$.
A prominent peak due to the ground state is observed
with a small contribution from the 2$^{+}_{1}$ state at $E_{x}=3.35$ MeV.
The resolutions of the excitation energy for the ground state
and the c.m. angle are 
$\Delta E_{x}=1$ MeV and $\Delta \theta_{\mathrm{c.m.}}=0.07^{\circ}$
in sigma, respectively.
The excitation-energy resolution is limited mainly due to the angular 
straggling of recoil alpha particles in the TPC gas.
As an example, angles of alpha particles at 2 MeV are straggled about 
30 mrad due to collisions with the gas particles at 1000 hPa.
By fitting the spectrum with two Gaussians, the yields of the ground
and 2$^{+}_{1}$ states were obtained.
At $\theta_{\mathrm{c.m.}}<5^{\circ}$, the yield of the $2_{1}^{+}$
state could not be determined because the contribution of the $2_{1}^{+}$ 
state was much smaller than that of the ground state.

The detection efficiency for the present measurement was estimated
by a Monte Carlo simulation.
It was assumed that the $\alpha+^{10}$C scattering occurs inside the
sensitive volume of the TPC over the entire solid angle.
Primary electrons were generated along the $^{10}$C and recoil alpha
trajectories according to the SRIM calculation,
considering the angular straggling of the recoil alpha particles.
These electrons drifted towards the $\mu$-PIC.
The transverse and longitudinal diffusions of the electrons were
taken into account.
The charge collection rate by the $\mu$-PIC as a function of time was folded
by the response function of the readout circuit to
simulate the analog signal from each strip.
The simulated signals were virtually processed and the track images
were generated.
These images were analyzed in the same manner as the real data.
The number of reconstructed events at each $\theta_{\mathrm{c.m.}}$ 
and $E_{x}$ was divided by the number of generated events to estimate 
the detection efficiency.
The efficiency depends on the recoil angle and energy in the laboratory
frame.
For example, when the recoil alpha particles are emitted to 
$\theta_{\mathrm{lab.}}=88^{\circ}$ with an energy of 0.8 MeV,
which corresponds to $\theta_{\mathrm{c.m.}}=4.5^{\circ}$ and $E_{x}=0$ MeV,
the efficiency for the alpha particles
to reach the trigger region, and to start the data acquisition
is about 36$\%$.
The track reconstruction efficiency for the recorded events
is about 82$\%$.
Consequently, the total efficiency is about 30$\%$.

\begin{figure}[tb]
\includegraphics[width=82mm]{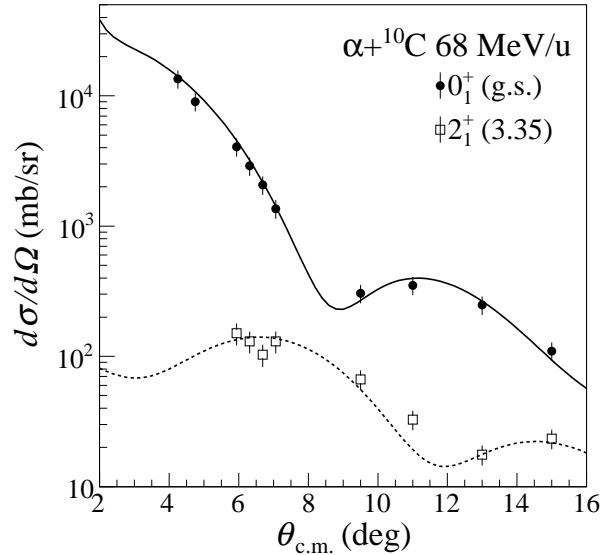}
\caption{
  Differential cross sections for the $\alpha+^{10}$C
  elastic (solid circles) and inelastic scatterings to the $2_{1}^{+}$ state
  at $E_{x}=3.35$ MeV (open squares).
  The cross section of the elastic scattering calculated with the 
  optical-model potential
  is shown by the solid line, while the cross section of the inelastic
  scattering obtained by the DWBA calculation is shown by the dashed line.
}
\label{10c_cross}
\end{figure}

\begin{figure}[tb]
\includegraphics[width=82mm]{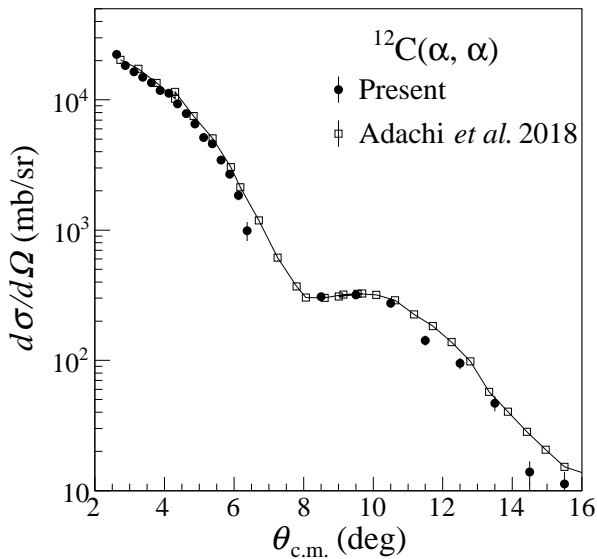}
\caption{
  Differential cross sections for the $\alpha+^{12}$C elastic
  scattering at 94 MeV/u (solid circles) compared with the previous
  results measured under the normal kinematic condition at
  96 MeV/u (open squares) \cite{adachi}.
  The solid line connecting the open squares is drawn to guide the eyes.}
\label{12c_cross}
\end{figure}

Finally, the differential cross sections of the $\alpha+^{10}$C elastic
scattering and the inelastic scattering exciting the $2_{1}^{+}$
state at $E_{x}=3.35$ MeV were obtained, as plotted in Fig. \ref{10c_cross}.
The cross section of the $\alpha+^{12}$C elastic scattering
is also obtained to check the present analysis.
In Fig. \ref{12c_cross}
the measured cross section is compared with the previous
result obtained using a $^{4}$He beam at 96 MeV/u  
under the normal kinematic condition \cite{adachi}.
The present result agrees with the previous result qualitatively, however,
it is systematically smaller than the previous result by 
10$\%$ on average.
The normalization factors of the present result to the previous 
result at different angles fluctuate $\pm16 \%$ around the averaged value.
This is mainly due to the uncertainty of the detection efficiency of MAIKo.
Therefore, we added 16$\%$ fractional uncertainty to the statistical 
uncertainty in quadrature in the following analysis.

\section{\label{sec-ana}Analysis}
We performed the DWBA calculation with single-folding potentials
to extract the
neutron transition matrix element $M_{n}$ in $^{10}$C from the
cross section of the alpha inelastic scattering to the $2_{1}^{+}$ state.
The phenomenological $\alpha$-$N$ effective interaction and
the ground-state density distribution were determined 
to reproduce the cross section of the alpha elastic scattering.
The DWBA calculations were performed using the computer code 
ECIS-95 \cite{ecis}.

\subsection{\label{sec-ana-a}Analysis of alpha elastic scattering}

The optical-model potential for the $\alpha+^{10}$C elastic scattering
was obtained by folding a phenomenological $\alpha$-$N$ effective interaction
$u$ with the point-nucleon distribution in the ground state $\rho(\bm{r'})$:
\begin{equation}
  U(r)=\int \rho(\bm{r'})u[|\bm{r}-\bm{r'}|,\ \rho(\bm{r'})] d\bm{r'},
  \label{eq_pot}
\end{equation}
where $\bm{r}$ and $\bm{r'}$ represent the positions of the 
alpha particle and nucleons in $^{10}$C, respectively.
The phenomenological $\alpha$-$N$ interaction was parametrized as given
in Ref. \cite{adachi}:
\begin{equation}
  \begin{split}
    u[|\bm{r}-\bm{r'}|,\ \rho(\bm{r'})]=
    -V&[1+\beta \rho^{2/3}(\bm{r'})]\mathrm{e}^{-|\bm{r}-\bm{r'}|^2/\alpha_{V}^2} \\
    -iW&[1+\beta \rho^{2/3}(\bm{r'})]\mathrm{e}^{-|\bm{r}-\bm{r'}|^2/\alpha_{W}^2}.
  \end{split}
  \label{eq_alphan}
\end{equation}
The parameters $V$ and $W$ are the depths of the real and imaginary potentials,
respectively.
$\beta$ is the density-dependent coefficient of the interaction.
$\alpha_{V}$ and $\alpha_{W}$ are the range parameters of the real and 
imaginary parts.
In the present analysis, we assumed that the real and imaginary ranges
are the same ($\alpha_{V}=\alpha_{W}=\alpha$), 
and the interaction is density-independent ($\beta=0$).
It was reported that the density-dependent ($\beta \neq 0$)
and density-independent interactions provide almost the same cross sections 
for the $2_{1}^{+}$ states in the DWBA calculation 
once the interaction parameters are determined 
to reproduce the cross section for the ground state \cite{adachi}.
Therefore, we chose the density-independent interaction for simplicity.

The point-nucleon distribution of the ground state 
in $^{10}$C was parametrized using the three-parameter Gaussian (3pG) function:
\begin{equation}
  \rho(\bm{r}) = 
  \frac{\rho_{0}(1+wr^{2}/c^{2})}{1+\mathrm{e}^{(r^{2}-c^{2})/z^{2}}}.
\end{equation}
Here, $c$, $z$, and $w$ are the parameters of the 3pG function.
The normalization factor $\rho_{0}$ is determined so as to satisfy the relation:
\begin{equation}
  \int \rho(\bm{r}) d\bm{r} = 4\pi\int \rho(r)r^2dr = A,
\end{equation}
where $A$ is the mass number.

In standard analysis, the density distribution of the ground state is 
taken from the electron elastic scattering, 
and only the effective interaction is optimized to 
reproduce the alpha elastic scattering. 
However, the density distribution in $^{10}$C is not known, 
and both the effective interaction and the density distribution 
must be determined simultaneously. 
Unfortunately, the effective interaction and the density distribution 
are not fully decoupled in the calculation of the cross section.
The range parameter $\alpha$ in the effective interaction 
and the radius of the density distribution are strongly coupled;
therefore, these parameters cannot be determined uniquely. 
In the present analysis, $\alpha$ was fixed at 2.13 fm.
This value was determined
by analyzing the $\alpha+^{12}$C elastic scattering
at 60 MeV/u \cite{12ca}
in the same manner as in Ref. \cite{adachi}.

The interaction parameters $V$ and $W$, and the 3pG parameters
$c$, $z$, and $w$ were optimized to reproduce the measured cross section
of the $\alpha+^{10}$C elastic scattering.
The obtained parameters are listed in Table \ref{tab_potden}, 
and the calculated cross section with these parameters 
is indicated by the solid line in Fig \ref{10c_cross}.
The reduced chi-square of the fit $\chi^{2}/ \nu$, 
where $\nu$ is the number of degrees of freedom, is 4.98/5.
The standard uncertainties of the parameters in the effective $\alpha$-$N$
interaction were estimated by varying one of the parameters over the range 
that satisfies the following relation:
\begin{equation}
  \chi^{2} \leq \chi^{2}_{\mathrm{min}}+1.
  \label{eq_error1}
\end{equation}
When the uncertainty of one parameter was estimated,
the other parameters were freely changed to minimize $\chi^2$.


\begin{table*}
\caption{Optimized parameters for the $\alpha$-$N$ effective interaction 
  and the point-nucleon distribution of the ground state in $^{10}$C 
  in the present analysis.}
\label{tab_potden}
\begin{ruledtabular}
\begin{tabular}{cccccccp{2in}}
  \multicolumn{3}{c}{Interaction}&\multicolumn{4}{c}{3pG}\\ \cmidrule(lr){1-3}  \cmidrule(lr){4-7}
  $\alpha$ (fm)&$V$ (MeV)&$W$ (MeV)& $c$ (fm)& $z$ (fm)& $w$ & rms (fm)\\ \hline
  $2.13$&$25.8_{-2.1}^{+3.1}$&$17.0_{-2.0}^{+2.7}$&$0.21$&$1.98$&$-1.8\times10^{-4}$&$2.6\pm0.3$
\end{tabular}
\end{ruledtabular}
\end{table*}

\begin{figure}[tb]
\includegraphics[width=82mm]{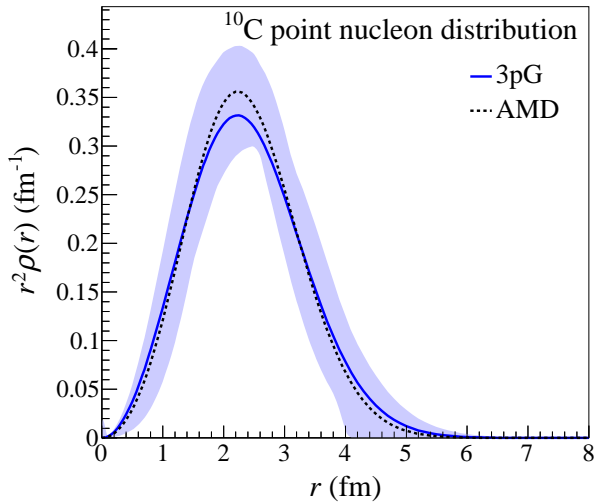}
\caption{Point-nucleon distribution of the ground state in $^{10}$C.
  The density distribution obtained from the present work is shown by the
  solid blue line associated with the error band.
  The dashed black line represents the AMD calculation \cite{enyo_amd}.
}
\label{10cden}
\end{figure}

The deduced point-nucleon distribution of the ground state in $^{10}$C is 
shown in Fig. \ref{10cden}.
The vertical axis represents $\rho(r)$ multiplied by $r^2$.
The distribution given by the best-fit parameters in 
Table \ref{tab_potden} is drawn by the solid blue line 
with its error band.
The error band was calculated by varying the three parameters 
in the 3pG function simultaneously over the range that satisfies
\begin{equation}
  \chi^{2} \leq \chi^{2}_{\mathrm{min}} + \Delta \chi^{2}.
  \label{eq_error2}
\end{equation}
When the error band was estimated, $V$ and $W$ in the
effective interaction were freely changed
to minimize $\chi^{2}$.
$\Delta \chi^{2}$ obeys the $\chi^{2}$ distribution 
for three degrees of freedom
since the 3pG function has three independent parameters \cite{stat}.
Thus, $\Delta \chi^{2}$ is equal to 3.53
at a confidence level of 68.3$\%$.
Since the effective interaction and the density distribution were 
simultaneously optimized to reproduce the measured cross section 
in the present analysis,
the deduced density distribution is associated with 
the large error band, as seen in Fig. \ref{10cden}.

The density distribution is compared with the theoretical one of the 
antisymmetrized molecular dynamics (AMD) model \cite{enyo_amd}
plotted by the black dashed line.
The present result is consistent with the AMD model.
The root-mean-square (rms) radius of the point-nucleon distribution
is $2.6\pm0.3$ fm, 
which is consistent with $2.42\pm0.10$ fm from the proton 
elastic scattering \cite{10cp} and 2.52 fm from the AMD calculation;
however, it is larger compared to $2.27\pm0.03$ 
fm, which is deduced from
the interaction cross section \cite{Ozawa1996}.

\subsection{\label{sec-ana-b}Analysis of alpha inelastic scattering}

The transition potential $\delta U(r)$ for the alpha inelastic scattering
to the $2^{+}_{1}$ state was obtained by folding the effective 
$\alpha$-$N$ interaction $u$
with the transition density $\delta \rho(r)$:
\begin{equation}
  \delta U(r) = \int \delta \rho(r)
  u[|\bm{r}-\bm{r'}|,\ \rho(\bm{r'})] d\bm{r'}.
  \label{eq_tp}
\end{equation}
The effective interaction was determined in Sec. \ref{sec-ana-a},
and the transition density between the ground state and 
the $2_{1}^{+}$ state 
was calculated by the macroscopic model \cite{td_satchler}:
\begin{equation}
  \delta \rho_{n(p)}(r) = -\delta_{n(p)} \frac{d}{dr}\rho_{n(p)}(r),
  \label{eq_td}
\end{equation}
where $\delta_{n(p)}$ is the deformation length for a neutron (proton)
and $\rho_{n(p)}(r)$ is the neutron (proton) density distribution 
in the ground state.
Assuming that the proton and neutron distributions have the same shape,
$\rho_p(r)=(Z/A)\rho(r)$ and $\rho_n(r)=(N/A)\rho(r)$ were used in the 
present analysis.

The transition matrix elements of neutron (proton) were calculated
from the transition densities using the formula:
\begin{equation}
  M_{n(p)} = \int r^{4} \delta \rho_{n(p)}(r) dr.
  \label{eq_strength}
\end{equation}
Since a proton is not a point-like particle in reality, 
its charge form factor should be taken into account 
when $M_{p}$ is compared with $B(E2;0_{1}^{+}\to2_{1}^{+})$.
However, the alteration due to the proton charge form factor 
is as low as a few percent,
and it is negligible compared to other uncertainties in the present analysis.
Once we assume a proton to be a point particle, the reduced electromagnetic
transition rate $B(E2;0_{1}^{+}\to2_{1}^{+})$ is related to $M_{p}$ by
Eq. (\ref{eq_be2}).
The proton deformation length $\delta_{p}$ was determined so as to 
reproduce the known $B(E2;0_{1}^{+}\to2_{1}^{+})$ value of
$44.0\pm1.5$ $e^{2}$fm$^{4}$ in $^{10}$C \cite{10c_be2}.
The neutron deformation length $\delta_{n}$ was determined to
reproduce the measured cross section of the inelastic alpha scattering
as shown by the dashed line in Fig. \ref{10c_cross}.
The result is $\delta_{n}=2.4$ fm.
From Eqs. (\ref{eq_td}) and (\ref{eq_strength}), this corresponds to
$M_{n}=6.9$ fm$^{2}$.
The reduced $\chi^{2}$ of the fitting is $\chi^{2}/ \nu = 10.9/7$.

The uncertainty of $M_{n}$ from the procedure to fit the cross 
section of the inelastic scattering is $\pm0.4$ fm$^{2}$.
The uncertainties of the interaction and 3pG parameters
also cause an additional uncertainty in $M_{n}$.
This uncertainty was estimated to be $\pm0.6$ fm$^{2}$
by propagating the uncertainty in those parameters into $M_{n}$.
The total uncertainty from the procedures to fit the experimental data
is $\pm0.7$ fm$^{2}$.

In Ref. \cite{adachi}, the transition matrix elements in stable 
self-conjugate even-even nuclei were obtained by analyzing cross sections from
alpha inelastic scattering on the basis of DWBA calculations 
with single-folding potentials in a way similar to the present work.
It was found that the matrix elements obtained by the inelastic alpha scattering
erratically differ from the electromagnetic 
transition matrix elements taken from $B(E2;0_{1}^{+}\to2_{1}^{+})$ 
with a standard deviation of $\pm 17 \%$.
Thus, we adopt 17$\%$ ($\pm1.2$ fm$^{2}$) as a systematic uncertainty due to 
the error in the DWBA analysis with single-folding potentials.

Finally, we obtained the $M_{n}$ value in $^{10}$C and its uncertainties as
\begin{equation}
  M_{n} = 6.9\, \pm0.7\, \mathrm{(fit)}\, \pm1.2\, \mathrm{(sys)}\,
  \mathrm{fm}^2.
\end{equation}

\section{\label{sec-dis}Discussion}
The present result of $M_n$ is larger than the previous result of
$M_{n}=5.51\pm1.09$ fm$^{2}$ determined by the proton inelastic
scattering \cite{10cp}.
This discrepancy between the present and previous results is
possibly because the authors in Ref. \cite{10cp} used the old
$B(E2;0_{1}^{+} \to 2_{1}^{+})$ value from Ref. \cite{be2_10c_old}.
This old value of $61.5\pm10$ $e^{2}$fm$^{4}$ is larger than 
the new value of $44.0\pm1.5$ $e^{2}$fm$^{4}$ reported 
in Ref. \cite{10c_be2}.
We took the $B(E2;0_{1}^{+} \to 2_{1}^{+})$ value from the recent
measurement because of the smaller uncertainty.
The larger $B(E2;0_{1}^{+} \to 2_{1}^{+})$ value
might lead to a smaller value of $M_{n}$.
If we take the old $B(E2;0_{1}^{+} \to 2_{1}^{+})$ value,
the $M_{n}$ value becomes 
$M_{n} = 5.7\, \pm0.6\, \mathrm{(fit)}\, \pm1.0\, \mathrm{(sys)}$ fm$^{2}$.
This value is consistent with the previous result.

Assuming charge symmetry in the $A=10$ system, $M_{n}$ in $^{10}$C
should be equal to $M_{p}$ in $^{10}$Be.
$M_{p}$ in $^{10}$Be is reported as $6.78\pm0.11$ fm$^{2}$ \cite{10be_be2}.
This value is actually close to the present $M_{n}$ value in $^{10}$C,
and thus, charge symmetry in the $A=10$ system is almost conserved.

The $M_{n}/M_{p}$ ratio in $^{10}$C deduced from the present measurement is
\begin{equation}
  M_{n}/M_{p} = 1.05\, \pm0.11\, \mathrm{(fit)}\, 
  \pm0.17\, \mathrm{(sys)}.
  \label{double_ratio}
\end{equation}
The transition from the ground state to the $2_{1}^{+}$ state
in $^{10}$C is almost isoscalar,
whereas a large $M_{n}/M_{p}$ ratio of $3.2\pm0.7$ was reported in $^{16}$C \cite{ong_18c}.
This indicates that the quadrupole transition in $^{10}$C is less neutron dominant
compared to that in $^{16}$C.
The $M_{n}/M_{p}$ value close to unity in $^{10}$C
shows that the effect of the $Z=6$ subshell closure is less evident
compared to the neutron-rich side.


The present results are compared with 
the theoretical predictions by the
$2p+2\alpha$ four-body cluster model \cite{Ogawa2000},
no-core shell model \cite{PhysRevC.66.024314},
shell model \cite{PhysRevC.70.054316},
Monte Carlo shell model \cite{PhysRevC.86.014302},
and AMD model \cite{enyo_prc_2011}
in Table \ref{tab_ine}.
In the cluster-model calculation, fully antisymmetrized ten-nucleon wave 
functions were built in a microscopic $2p+2\alpha$ configuration space
using the Minnesota interaction \cite{THOMPSON197753}.
The no-core shell model calculation was performed
using the CD-Bonn $NN$ potential in the basis space up to $8\hbar\Omega$
with the harmonic oscillator frequency of $\hbar\Omega=14$ MeV.
The shell-model calculation was conducted
within the $p$ shell using the Cohen and Kurath (8--16)2BME interaction
\cite{COHEN19651}. 
The shell-model transition matrix elements were calculated 
using single-particle wave 
functions in the harmonic oscillator potential with $b=1.64$ fm
and effective charges of $e_{p}=1.3e$ and $e_{n}=0.5e$.
In the Monte Carlo shell-model calculation,
the unitary correlation operator method potential based on
the chiral next-to-next-to-next-to-leading-order two nucleon
interaction \cite{Machleidt2011} with
the bare charges ($e_{p}=e$ and $e_{n}=0$) was used to calculate
the transition matrix elements.
The $M_{n}$ value in $^{10}$C from the Monte Carlo shell-model calculation
was taken from the $M_{p}$ value
in the mirror nucleus $^{10}$Be assuming charge symmetry.

Theoretical calculations systematically underestimate $M_{p}$ and $M_{n}$ 
except for the Monte Carlo shell-model calculation 
and $M_{n}$ calculated by the AMD model.
Especially, $M_{p}$ and $M_{n}$ predicted by the shell-model calculation 
are considerably smaller than the experiment. 
However, the $M_{n}/M_{p}$ ratios from the theoretical calculations are
close to unity, supporting the present result that enhancement of the
$M_{n}/M_{p}$, which was observed in neutron-rich $^{16}$C,
was not observed in proton-rich $^{10}$C.

\begin{table*}
  \caption{
    Experimental transition matrix elements of proton and neutron 
    from the ground state
    to the $2_{1}^{+}$ state in $^{10}$C compared with theoretical calculations.
    The $M_{n}/M_{p}$ ratios are also listed.
  }
  \label{tab_ine}
  \begin{ruledtabular}
    \begin{tabular}{ccccp{2in}}
      &$M_{p}$ (fm$^{2}$)&$M_{n}$ (fm$^{2}$)&$M_{n}/M_{p}$\\ \hline
      Experiment&$6.63\pm0.11$ \cite{10c_be2}&$6.9\, \pm0.7\, \pm1.2$&$1.05\, \pm0.11\, \pm0.17$\\
      Cluster \cite{Ogawa2000}&$5.5$&$4.4$&$0.8$\\
      No-core shell model \cite{PhysRevC.66.024314}&$5.3$&$5.7$&$1.1$\\
      Shell model \cite{PhysRevC.70.054316}&$3.3$&$4.3$&$1.3$\\
      Monte Carlo shell model \cite{PhysRevC.86.014302}&$6.8$&$6.8$&$1.0$\\
      AMD \cite{enyo_prc_2011}&$5.3$&$6.9$&$1.3$
    \end{tabular}
  \end{ruledtabular}
\end{table*}

\section{\label{sec-sum}Summary}
Alpha elastic and inelastic scatterings from $^{10}$C at 68 MeV/u were
measured under the inverse kinematic condition at RCNP, Osaka University.
The purity and intensity of the $^{10}$C secondary beam were 96$\%$ and
70 kcps, respectively.
The recoil alpha particles were detected using the newly developed 
MAIKo active target system \cite{maiko}.
This system enabled the detection of low-energy recoil alpha particles 
down to $E_{\alpha}=0.5$ MeV,
which corresponds to momentum transfer down to
$q=0.4$ fm$^{-1}$.
The excitation-energy resolution was approximately 1 MeV in sigma, 
sufficient to distinguish
the first excited $2_{1}^{+}$ state at $E_{x}=3.35$ MeV from the 
ground state in $^{10}$C.

The cross section for the $\alpha+^{12}$C elastic scattering was also 
measured using a primary $^{12}$C beam at 94 MeV/u.
The measured cross section was compared with the previous result obtained 
under the normal kinematic condition
using a $^{4}$He beam at 96 MeV/u \cite{adachi},
and we confirmed that both results are qualitatively consistent.

The cross section of the $\alpha+^{10}$C
elastic scattering enabled the determination of 
the phenomenological $\alpha$-$N$
effective interaction and
the point-nucleon distribution of the ground state in $^{10}$C.
The rms radius of $2.6\pm0.3$ fm in $^{10}$C
is consistent with the theoretical prediction by the AMD calculation \cite{enyo_amd}
and the experimental result of the previous proton elastic scattering \cite{10cp},
but slightly larger than that deduced from the interaction cross section \cite{Ozawa1996}.

From the cross section of the $\alpha+^{10}$C inelastic scattering to the 
$2^{+}_{1}$ state,
the neutron transition matrix element of 
$M_{n} = 6.9\, \pm0.7\, \mathrm{(fit)}\, \pm1.2\, \mathrm{(sys)}$
was obtained.
The $M_{n}/M_{p}$ ratio in $^{10}$C was determined as
$M_{n}/M_{p} = 1.05\, \pm0.11\, \mathrm{(fit)}\, \pm0.17\, \mathrm{(sys)}$,
and thus, the effect of the $Z=6$ subshell closure reported in neutron-rich
carbon isotopes \cite{z6magic} is not evident in the proton-rich side.
This result is supported from the theoretical calculations.

The first physics experiment using the MAIKo active target was 
successfully completed.
MAIKo will be employed in various RI beam experiments in the near future.

\begin{acknowledgments}
  The authors are grateful to the RCNP cyclotron crews
  for the stable operation of the cyclotron facilities.
  Discussions with Prof. N. Aoi, Prof. M. Ito, Prof. K. Ogata, and Prof. H. Sakaguchi
  were of great help in the experimental planning and the analyses.
  T.~F. appreciates the support of Grant-in-Aid for JSPS Research 
  Fellow JP14J00949.
  This research was supported by JSPS KAKENHI,
  Grants No. JP20244030, No. JP23340068, 
  No. JP23224008, No. JP15H02091, and No. JP19H05153.
\end{acknowledgments}


\bibliography{prc}

\end{document}